\documentclass[a4paper]{jpconf}
\usepackage{xcolor}
\usepackage{graphicx}
\usepackage{amsmath}
\begin{document}
\title{On the Use of the Tsallis Distribution at  LHC Energies.}

\author{J.~Cleymans}
\address{UCT-CERN Research Centre and Physics Department, University of Cape Town, South Africa}
\ead{jean.cleymans@uct.ac.za}
\begin{abstract}
Numerous papers have appeared recently showing fits to
transverse momentum ($p_T$) spectra measured at the Large Hadron Collider (LHC) in proton - proton 
collisions.This talk focuses on the fits extending to very large values of the transverse momentum  with 
$p_T$ values up to 200 GeV/c as measured by the ATLAS and CMS collaborations
at $\sqrt{s}$ = 0.9 and 7 TeV. 
A thermodynamically consistent form of the 
Tsallis distribution is used for fitting the transverse momentum spectra at mid-rapidity. The fits based on 
the proposed distribution provide an excellent description over 14 orders of magnitude.
Despite this success, an ambiguity is noted concerning the determination of the 
parameters in the Tsallis distribution.
This prevents drawing firm conclusions  as to the 
universality of the parameters appearing in the Tsallis distribution.
\end{abstract}

It was shown in numerous recent publications that fits based on the Tsallis distribution~\cite{tsallis}
give a good description of transverse momentum distributions measured at the 
LHC~\cite{biro,zheng1,gao,zheng2,zheng3,wilk,marques,urmossy,sorin}.  
Some of these fits  extend 
to values of $p_T$ up to 200 GeV/c~\cite{wong1,wong2,wilk2,azmi,azmi-cleymans} and provide an excellent description over 14 
orders of magnitude in the transverse momentum spectrum.
 
Of the many forms proposed, one in particular~\cite{worku1,worku2} leads to
a consistent version  of
thermodynamics for 
the particle number, energy density and pressure. In particular, the momentum distribution of particles is given by 
the following expression 
\begin{equation}
E\frac{dN}{d^3p} = gVE\frac{1}{(2\pi)^3}
\left[1+(q-1)\frac{E-\mu}{T}\right]^{-\frac{q}{q-1}} .
\end{equation}
where $g$ is the degeneracy factor, $E$ the energy, $p$ the momentum,  $V$ the volume of the system and $\mu$ is the chemical
potential.
The advantages are that all  thermodynamic consistency conditions are satisfied:
$$N = V\left.\frac{\partial P}{\partial \mu}\right|_{T,V}~~~~~\text{etc...} ,$$
and the parameter $T$ truly deserves its name since 
$$ T = \left.\frac{\partial E}{\partial S}\right|_{V,N}  .$$
In terms of the rapidity and transverse mass variables, 
$E = m_T\cosh y$, 
Eq. (1) becomes
(at mid-rapidity $y=0$ and  for $\mu$ = 0)
\begin{equation}
\left.\frac{d^2N}{dp_T~dy}\right|_{y=0} 
= gV\frac{p_Tm_T}{(2\pi)^2}
\left[ 1+(q-1)\frac{m_T}{T} \right]^{-\frac{q}{q-1}} .
\end{equation}
%
%
Integrating the above expression over the rapidity variable leads to the equivalent form first derived
in~\cite{ryb}
\begin{eqnarray}
\left.\frac{d^2N}{dp_T~dy}\right|_{y=0} 
&=& \left.\frac{dN}{dy}\right|_{y=0}   \frac{p_T m_T}{T} \left[1+(q-1)\frac{m_T}{T}\right]^{-q/(q-1)} \nonumber\\
&&\times  \frac{(2-q)(3-2 q)}{(2-q)m^2 + 2 m T + 2   T^2} \nonumber\\
&&\times  \left[1+(q-1)\frac{m}{T}  \right]^{1/(q-1)}     
\end{eqnarray}
or, showing only the dependence on the transverse variables:
\begin{eqnarray}
\left.\frac{d^2N}{dp_T~dy}\right|_{y=0} 
&=&  \left.\frac{dN}{dy}\right|_{y=0} p_T \frac{m_T}{T}\left[1+(q-1)\frac{m_T}{T}\right]^{ -q/(q-1)}\nonumber\\
&&\times (\text{factors independent of}~p_T)
\end{eqnarray}
At large transverse momenta  the asymptotic behavior  is 
\begin{equation}
\lim_{p_T\rightarrow\infty}\left.\frac{d^2N}{dp_T~dy}\right|_{y=0} 
\propto  p_T \left[\frac{p_T}{T}\right]^{ -1/(q-1)}
\nonumber
\end{equation}
which shows that the scale is being set by the temperature $T$ and the asymptotic behavior is set by $q$. It is
highly sensitive to small deviations from 1 in this last variable.  The upper limit for $q$ is given by
\begin{equation}
q < \frac{4}{3}  .
\end{equation}
For larger values of $q$ the integrals become divergent~\cite{mogliacci}.

The above  power law has been used to fit the $p_T$ spectra of charged particles measured by the 
ATLAS~\cite{atlas} and CMS~\cite{CMS} collaborations in~\cite{azmi}.  
The ATLAS collaboration has reported the transverse momentum in an inclusive phase space 
region taking into account at least two charged particles in the kinematic range $|\eta| < 2.5$ and $p_T > 100$ MeV~\cite{atlas}. 
The CMS collaboration has presented the differential transverse momentum distribution covering a $p_T$ range up to 200 GeV/c, the 
largest range ever measured in a colliding beam experiment~\cite{CMS}. The fits are presented in Figs. (1) and (2).

The results can be compared to those obtained in~\cite{wilk,wong1,wong2} where very good fits to 
transverse momentum distributions were presented. The quality of the fits s confirmed albeit with  
different values of the  parameters since a  different version is being used.
\begin{figure}[h]
\begin{minipage}{17pc}
\includegraphics[width=17pc]{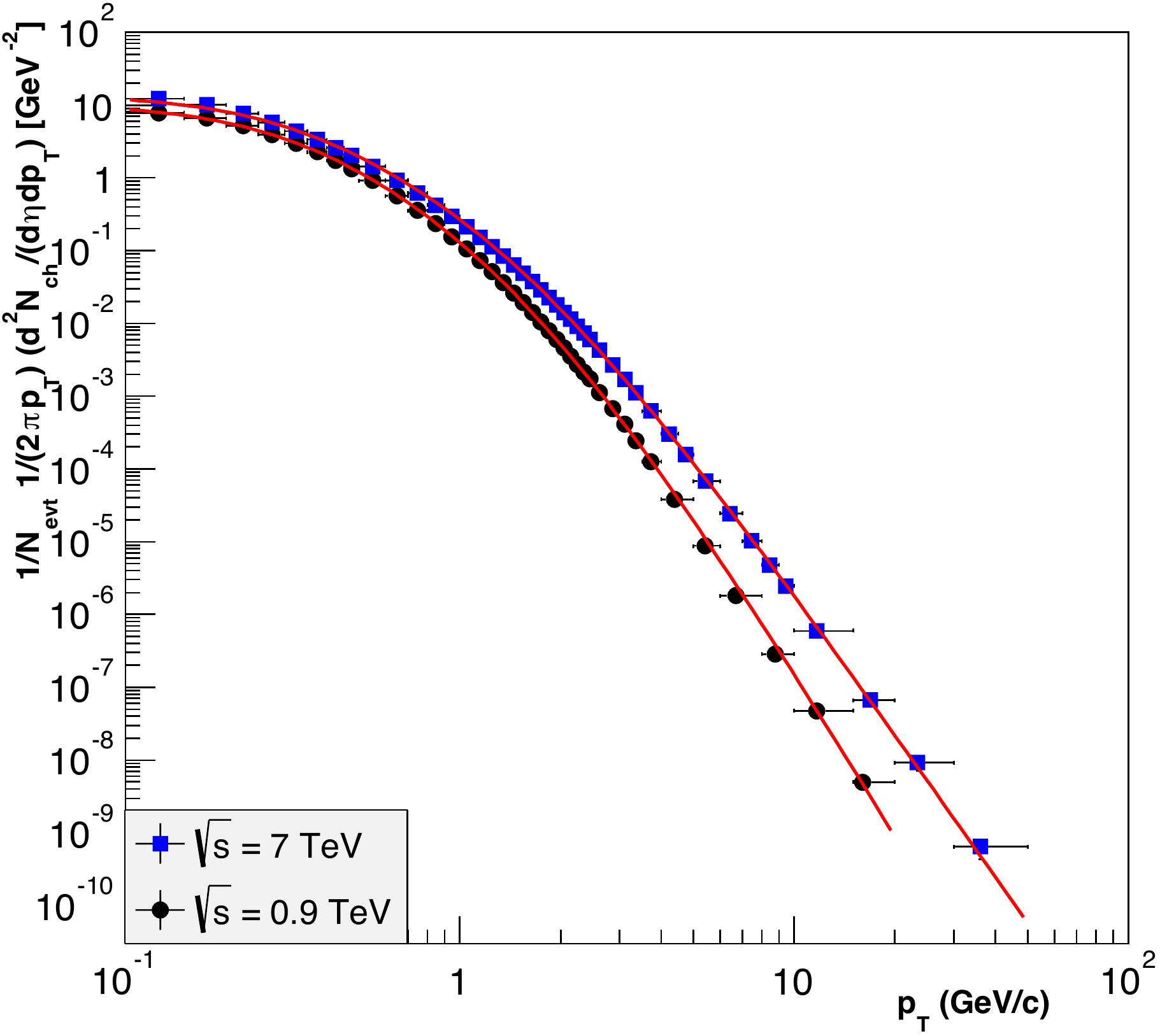}
\caption{Charged particle multiplicities as a function of the transverse momentum measured by the ATLAS collaboration 
for events with $n_{ch} \ge 2$, $p_T > $100 MeV and $|\eta| < $ 2.5 at $\sqrt{s}$ = 0.9 and 7 TeV in 
proton - proton collisions~\cite{atlas} fitted with Tsallis distribution~\cite{azmi}.}
\end{minipage}\hspace{2pc}%
\begin{minipage}{17pc}
\includegraphics[width=17pc]{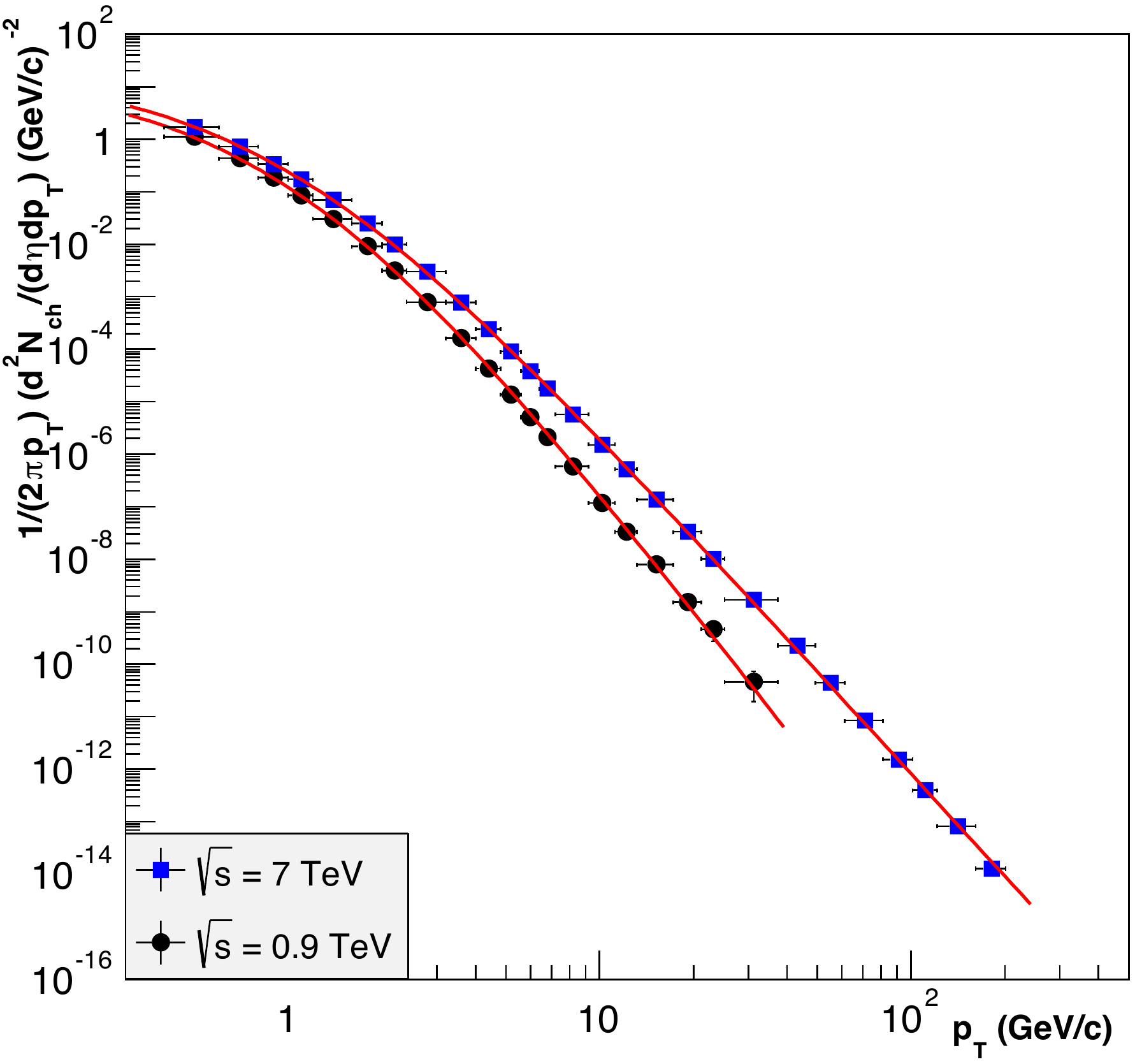}
\caption{Charged particle differential transverse momentum yields measured within $|\eta| < 2.4$ by the 
CMS collaboration in proton - proton collisions at $\sqrt{s}$ = 0.9 and 7 TeV~\cite{CMS} fitted with 
the Tsallis distribution~\cite{azmi}.}
\end{minipage} 
\end{figure}

The resulting parameters are listed in table 1.
\begin{table}
\begin{tabular}{lccccc}
\hline
{\bf Experiment} & \multicolumn{1}{c}{\bf $\sqrt{s}$ (TeV)} & \multicolumn{1}{c}{\bf $q$} & \multicolumn{1}{c}{\bf $T$ (MeV)} & \multicolumn{1}{c}{\bf $R$ (fm)} & \multicolumn{1}{c}{\bf $\chi^2/NDF$} \\
\\
\hline
ATLAS & 0.9 & 1.129 $\pm$ 0.005 & 74.21 $\pm$ 3.55 & 4.62 $\pm$ 0.29 & 0.657503/36 \\
ATLAS & 7 & 1.150 $\pm$ 0.002 & 75.00 $\pm$ 3.21 & 5.05 $\pm$ 0.07 & 4.35145/41 \\
\hline
CMS & 0.9 & 1.129 $\pm$ 0.003 & 76.00 $\pm$ 0.17 & 4.32 $\pm$ 0.29 & 0.648806/17 \\
CMS & 7 & 1.153 $\pm$ 0.002 & 73.00 $\pm$ 1.42 & 5.04 $\pm$ 0.27 & 0.521746/24 \\
\hline
\end{tabular}
\caption{Values of the $q$, $T$ and $R$ parameters and $\chi^2/NDF$ obtained from fits to the $p_T$ spectra measured by the ATLAS~~\cite{atlas} and CMS~\cite{CMS} collaborations.}
\label{par1}
\end{table}
It is quite remarkable that the  transverse momentum distributions measured up to 200 GeV/c in $p_T$ can be described 
consistently over 14 orders of magnitude by a straightforward Tsallis distribution. \\
Despite this success, an ambiguity has been noted concerning the determination of the 
parameters~\cite{barnby} in the Tsallis distribution.
This is shown explicitly  in Fig.~(3) where fits are presented to the distributions of proton  measured by the 
CMS collaboration~\cite{CMSppbar}. 
In the figure, as an example,  three different sets of variables are shown  reproducing the measurements. This
shows clearly that a larger interval in the transverse momentum is needed before the parameters can be  determined accurately.
The solid black line corresponds to a temperature $T = 29.6$  MeV, the dashed line has $T = 52.3$ MeV and the dash-dotted
line has $T = 73.00$ MeV as determined from the fit to charged particles at the same energy. The first two choices for $T$ and the corresponding changes in $q$ and $dN/dy$   give equally acceptable fits despite the large change especially in the temperature $T$.
Thus  no firm conclusion can be drawn at present as to the 
universality of the parameters appearing in the Tsallis distribution as given above.
%
\begin{figure}
\begin{center}
\includegraphics[width=0.9\linewidth,height=12cm]{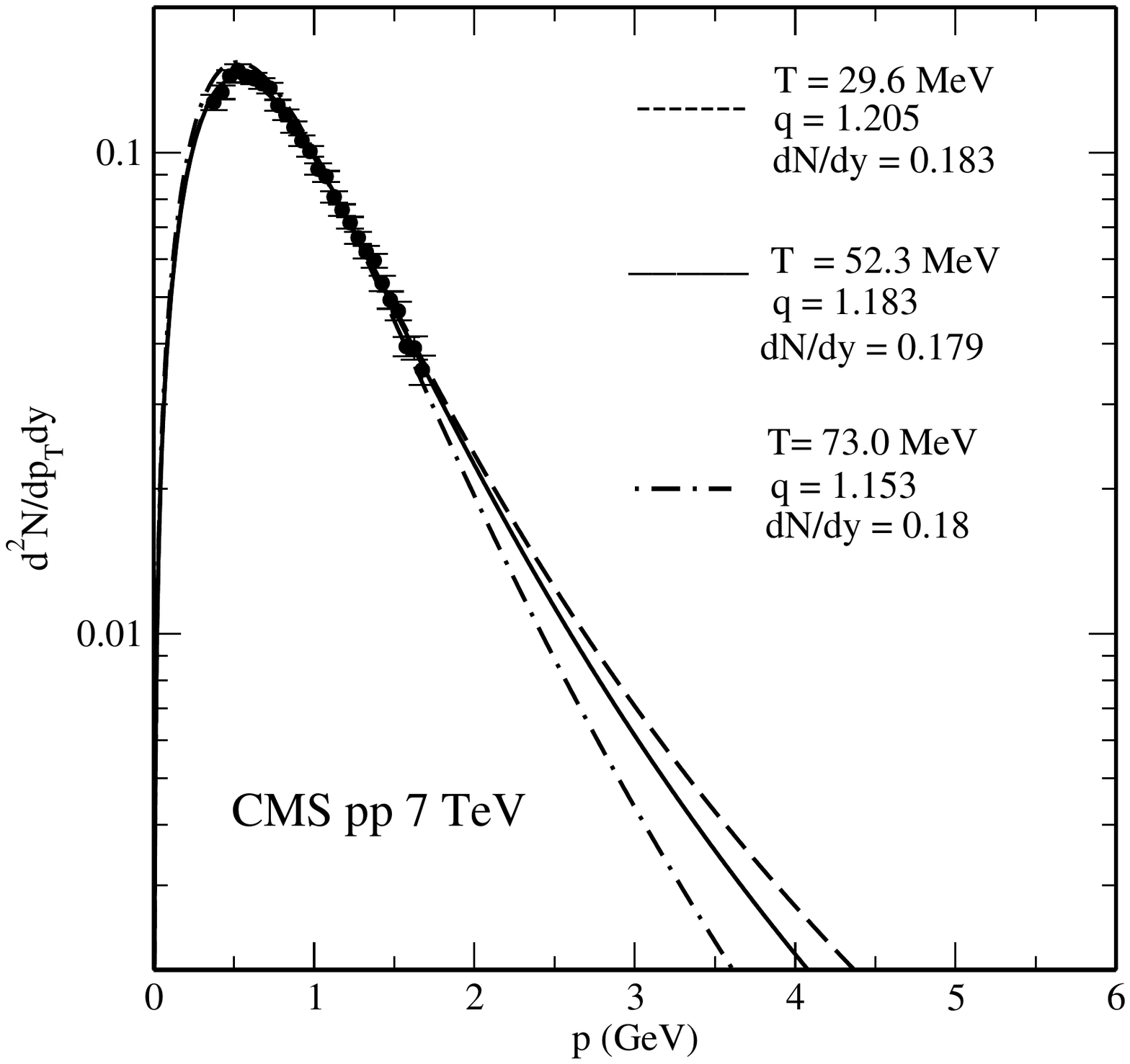}
\end{center}
\caption{Distribution of protons measured by the CMS collaboration~\cite{CMSppbar} in $p-p$ collisions
at 7 TeV. Three different Tsallis fits are shown with the parameters indicated in the figure.
}
\end{figure}

%
\end{document}